\documentclass[conference]{IEEEtran}
\pdfoutput=1
\IEEEoverridecommandlockouts
\usepackage{subfigure}
\usepackage{cite}
 \usepackage{siunitx}
\usepackage{amsmath,amssymb,amsfonts}
\usepackage{algorithmic}
\usepackage{algorithm}
\usepackage{graphicx}
\usepackage{textcomp}
\usepackage{xcolor}
\usepackage{listings}
\usepackage{booktabs} 
\usepackage{stfloats}
\usepackage{balance}

\def\BibTeX{{\rm B\kern-.05em{\sc i\kern-.025em b}\kern-.08em
    T\kern-.1667em\lower.7ex\hbox{E}\kern-.125emX}}
\begin{document}

\title{MTCA: Multi-Task Channel Analysis \\ for Wireless Communication}

\author{Jun Jiang\IEEEauthorrefmark{1}, Wenjun Yu\IEEEauthorrefmark{1}, Yuan Gao\IEEEauthorrefmark{1}\IEEEauthorrefmark{3}, and Shugong Xu\IEEEauthorrefmark{2}\IEEEauthorrefmark{3}\\
\IEEEauthorrefmark{1} School of Communication and Information Engineering, Shanghai University, Shanghai, China \\
\IEEEauthorrefmark{2} Xi'an Jiaotong-Liverpool University, Jiangsu, China \\
Email: \{jun\_jiang, yuwenjun, gaoyuansie\}@shu.edu.cn, Shugong.Xu@xjtlu.edu.cn \\
\IEEEauthorrefmark{3} Corresponding author
}

\maketitle

\begin{abstract}
In modern wireless communication systems, the effective processing of Channel State Information (CSI) is crucial for enhancing communication quality and reliability. However, current methods often handle different tasks in isolation, thereby neglecting the synergies among various tasks and leading to extract CSI features inadequately for subsequent analysis. To address these limitations, this paper introduces a novel Multi-Task Channel Analysis framework named MTCA, aimed at improving the performance of wireless communication even sensing. MTCA is designed to handle four critical tasks, including channel prediction, antenna-domain channel extrapolation, channel identification, and scenario classification. Experiments conducted on a multi-scenario, multi-antenna dataset tailored for UAV-based communications demonstrate that the proposed MTCA exhibits superior comprehension of CSI, achieving enhanced performance across all evaluated tasks. Notably, MTCA reached 100\% prediction accuracy in channel identification and scenario classification. Compared to the previous state-of-the-art methods, MTCA improved channel prediction performance by 20.1\% and antenna-domain extrapolation performance by 54.5\%.
\end{abstract}


\section{Introduction}

The large-scale multiple input multiple output (MIMO) system, as a key technology in modern wireless communications, significantly enhances system capacity and transmission rates. With the development of large-scale MIMO systems, the size of antenna arrays has increased significantly, making it both costly and impractical to acquire complete downlink channel state information (CSI) from all antenna elements. Therefore, antenna-domain channel extrapolation is an effective strategy to acquire partial CSI and predict the channel response across the entire antenna array. This approach\cite{antenna_extra1,antenna_extra2} can greatly reduce the complexity of channel estimation and enhance the overall system performance.

In high-mobility scenarios, such as high-speed trains, unmanned aerial vehicles (UAVs), and vehicular networks, the phenomenon of channel aging becomes particularly pronounced\cite{truong2013effects}. To tackle this issue, channel prediction  techniques\cite{kim2020massive,wu2021channel,jiang2019neural,seq2seq} have emerged. Based on historical CSI to predict future channel states, thereby reducing the costs and resource consumption associated with frequent CSI updates.

Meanwhile, channel identification\cite{nlos1} is an indispensable component in wireless communications. It involves distinguishing between line of sight (LOS) and non-LOS (NLOS) conditions, which is crucial for traditional positioning algorithms based on azimuth angle of arrival (AoA) or time of arrival (ToA). Accurate channel identification enables the system to better adapt to different propagation environments, thus improving positioning accuracy and service quality.

\begin{figure}[tbp]
\centering\includegraphics[width=0.5\textwidth]{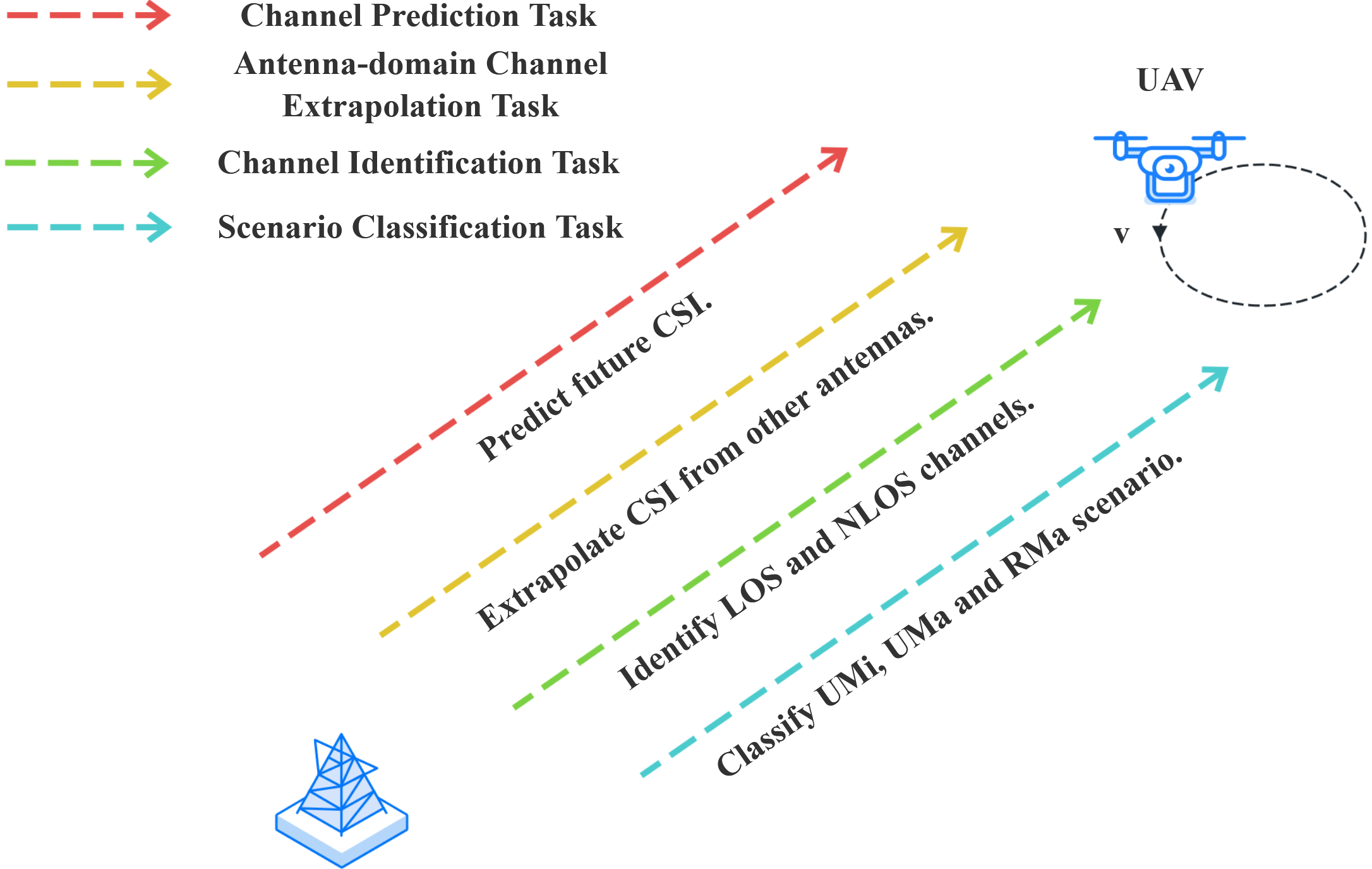}
\caption{Schematic diagram of four downlink tasks.}
\label{tasks}
\end{figure}

Furthermore, scenario classification plays a pivotal role in modern wireless communication systems. With advances in artificial intelligence (AI) technologies, AI-based channel modeling have garnered increasing attention. These methods rely on accurate classification of specific application scenarios because channel characteristics can vary significantly in different settings, such as Urban Macro (UMa), Urban Micro (UMi),
and Rural Macro (RMa). Scenario classification helps optimize the selection of channel models, thereby enhancing system performance across a variety of environments\cite{sce1} .

As shown in Fig~\ref{tasks}, channel prediction, antenna-domain channel extrapolation, channel identification, and scenario classification are all indispensable techniques in modern wireless communication systems. While they address distinct issues, there is a close relationship among them, collectively impacting the overall performance of wireless communication systems. Existing AI techniques, especially multi-task learning (MTL), have shown great potential in handling these tasks more efficiently and accurately by leveraging the correlations among them. MTL is a machine learning paradigm that trains a model on multiple related tasks simultaneously, improving its performance on individual tasks by sharing representations and learning common features \cite{mtlsurvey}.

Several approaches to MTL have been proposed, including hard parameter sharing and soft parameter sharing. Hard parameter sharing \cite{hard} involves sharing hidden layers between different tasks while allowing for task-specific output layers, which significantly reduces the risk of overfitting. Soft parameter sharing \cite{soft} allows each task to have its own model with its parameters, but it regularizes these models to be similar, offering more flexibility.

In the context of wireless communication systems, MTL can be applied to perform channel prediction, antenna-domain channel extrapolation, channel identification, and scenario classification concurrently. Therefore, we propose a novel architecture for Multi-Task Channel Analysis, named MTCA, which employs hard parameter sharing. By training a single model on these tasks, the model can learn shared features, leading to better performance than training separate models on each task independently. This approach can enhance the overall performance of wireless communication systems by leveraging the commonalities and differences among the tasks. 

The main contributions of this paper can be summarized as follows.
\begin{enumerate}
    \item We propose a novel Multi-Task Channel Analysis (MTCA) framework, designed to enhance the performance of wireless communication. MTCA integrates four critical tasks: channel prediction, antenna-domain channel extrapolation, channel identification, and scenario classification. To the best of our knowledge, MTCA represents the first attempt to concurrently address these four critical tasks within a unified model.
    
    \item We have developed a 5G channel dataset that complies with the 3GPP TR 36.777 standard and is tailored for UAV ground-to-air communications. This dataset covers UMa, UMi, and RMa scenarios, including both LOS and NLOS conditions. The dataset serves as an essential resource for model evaluation within this study and will be made accessible upon request to the corresponding author, thereby facilitating research reproducibility.
    
    \item Experiments using the aforementioned dataset demonstrate that MTCA achieves significant performance improvements over supervised single-task models, while maintaining a comparable model parameter. Specifically, MTCA achieves 100\% accuracy for channel identification and scenario classification. Compared with previous works, MTCA attains state-of-the-art performance, demonstrating a 20.1\% enhancement in channel prediction and a 54.5\% improvement in antenna-domain extrapolation compared to existing approaches.
    
\end{enumerate}

\section{System model}
Consider a multi-input single-output (MISO) UAV ground-to-air communication system, where the base station is the transmitter $ T_x $, and the UAV is the receiver $ R_x $. The speed of the UAV is denoted by $ v_{R_x} $. The base station is equipped with $ N_a $ transmitting antennas, and the receiving antenna of the UAV is modeled as an ideal point source. A cluster-based multipath channel model \cite{38_901} is adopted to represent the downlink CSI between the base station and the UAV. For the $ n_a $-th transmit-receive antenna pair, at time $ t $ and frequency $ f $, the channel response $ h_{n_a}(t, f) $ can be expressed as:

\begin{align}
    h_{n_a}(t, f) &= \sum_{n=1}^{N} \sum_{m=1}^{M_n} \alpha_{n,m} e^{j \left[ 2\pi \left( \nu_{n,m} t - f \tau_{n,m} \right) + \Phi_{n,m} \right]} \notag \\
    &\quad \times a_{T_x}\left( \phi_{m,n,n_a}(t), \theta_{m,n,n_a}(t) \right)
\end{align}

where $ N $ represents the total number of clusters in multipath propagation, and $ M_n $ is the number of sub-paths in the $ n $-th path. $ \alpha_{n,m} $ is the complex fading coefficient of the $ m $-th sub-path in the $ n $-th path, encapsulating information about path loss and phase shift. $ \nu_{n,m} $, $ \tau_{n,m} $, and $ \Phi_{n,m} $ are the Doppler shift, propagation delay, and random initial phase of the $ m $-th sub-path in the $ n $-th path, respectively. $ a_{T_x} \left( \phi_{m,n,n_a}(t), \theta_{m,n,n_a}(t) \right) $ is the directional weighting function of the transmitting antenna array, depending on the azimuth angle of departure $ \phi_{m,n,n_a}(t) $ and the elevation angle of departure $ \theta_{m,n,n_a}(t) $.

The Doppler shift $ \nu_{n,m} $ in the above equation can be expanded as:

\begin{equation}
    \nu_{n,m} = \frac{v_{R_x} \cdot \cos(\gamma_{n,m}) \cdot f}{c}
\end{equation}

where $ v_{R_x} $ is the speed of the UAV, $ \gamma_{n,m} $ is the angle between the UAV's velocity vector and the propagation path, and $ c $ is the speed of light.

In this system, the transmitter base station $ T_x $ is modeled as a Uniform Planar Array (UPA). The response vector of the transmitting antenna array $ a_{T_x}(\phi, \theta) $, corresponding to the spatial position and orientation of each antenna element, can be abstractly expressed as:
\begin{small}
\begin{equation}
a_{T_x}(\phi, \theta) = \left[ 1, e^{j 2\pi d \sin(\theta)\cos(\phi)}, \dots, e^{j 2\pi (N_a - 1) d \sin(\theta)\cos(\phi)} \right]^T
\end{equation}
\end{small}

where $ d $ is the spacing between antenna elements, and $ \phi $ and $ \theta $ represent the azimuth angle of departure and the elevation angle of departure, respectively. In this configuration, each antenna element is assumed to have no spatial rotation, with a rotation angle of $ 0^\circ $ along the $ x $-axis, $ y $-axis, and $ z $-axis.

For the tasks of scenario classification and channel identification, it is imperative to consider the diversity and particularities of various scenarios and propagation conditions. The 3GPP TR38.901 \cite{38_901} delineates five propagation scenarios: UMa, UMi, RMa, Indoor Office, and Indoor Factory. Each scenario is characterized by specific parameters that represent unique communication conditions and environmental features.

In the UMa scenario, the base station antennas are generally mounted at an elevation of 25 meters, mimicking a dense urban setting where the antennas are situated above the surrounding structures, thereby extending the coverage area, with a cell radius estimated around 500 meters. 

The UMi scenario involves base station antennas positioned at a height of approximately 10 meters, appropriate for densely populated urban areas with smaller coverage radii. Here, signals are prone to obstruction by buildings, leading to propagation paths dominated by blockage, diffraction, and scattering phenomena, with a cell radius typically measuring 200 meters.

The RMa scenario reflects aerial communications in suburban or rural locales, where the base station antennas are placed at a height of 35 meters. Given the unobstructed environment, these base stations command a broader coverage area, with inter-site distances potentially reaching up to 1732 meters or even 5000 meters.

Propagation conditions encompass LOS and NLOS environments. Under LOS conditions, the signal traverses directly between the transmitter and receiver, generally achieving superior transmission performance and signal integrity. Conversely, under NLOS conditions, the signal encounters interference through reflection, scattering, or diffraction from physical obstructions, engendering a more intricate propagation path that may induce multipath effects and signal degradation, thereby compromising transmission efficacy and quality.

\section{Propose framework}
\begin{figure}[tbp]
\centering\includegraphics[width=0.4\textwidth]{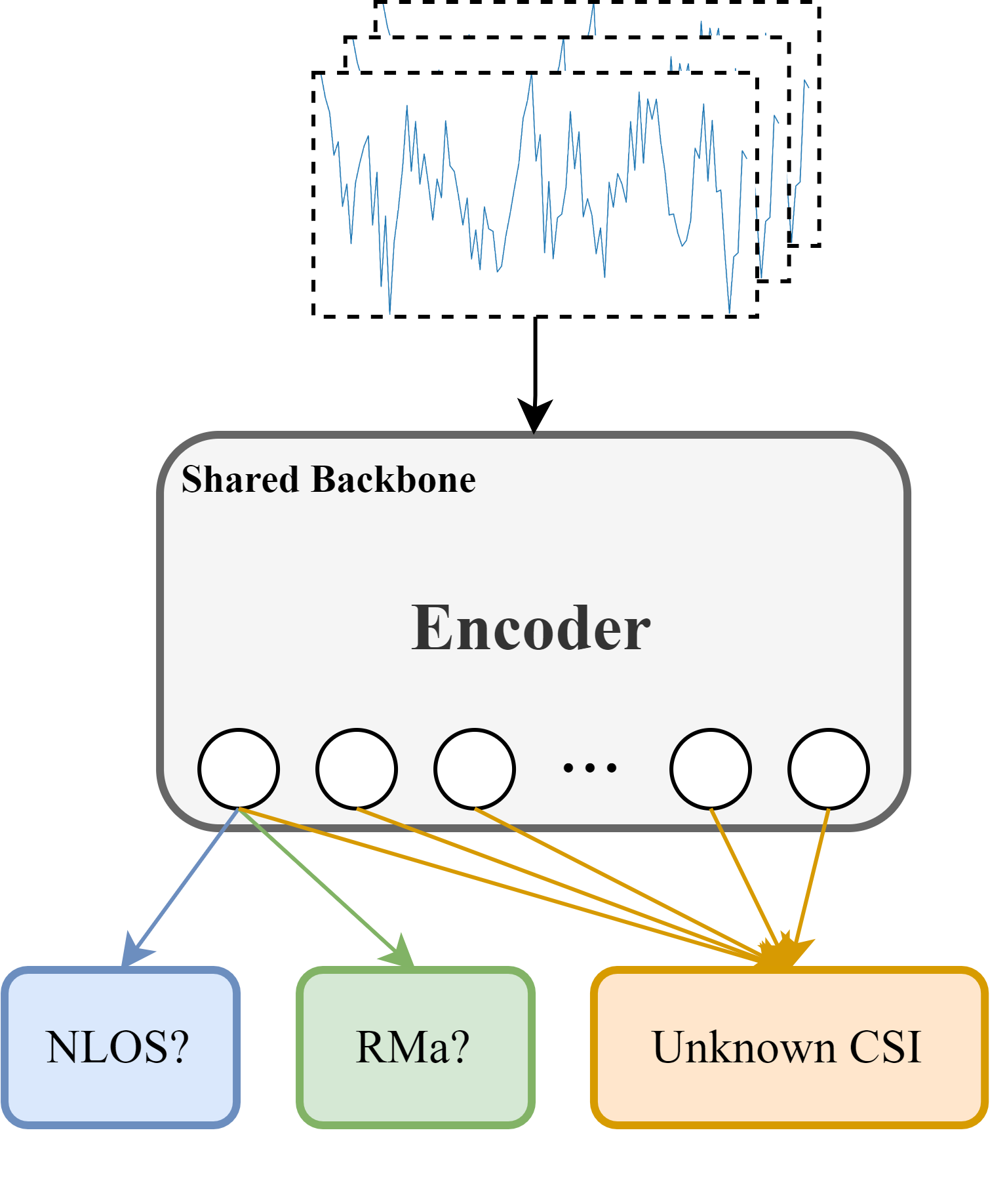}
\caption{Architecture of the proposed MTCA.}
\label{pipeline}
\end{figure}

In this section, we present a novel multi-task channel analysis framework, which leverages a shared gate recurrent unit (GRU) \cite{stenhammar2024comparison} as the encoder network coupled with multiple output heads based on fully connected layers (FC), termed MTCA (illustrated in Fig. \ref{pipeline}). The MTCA is specifically designed to concurrently achieve channel prediction, antenna-domain channel extrapolation, channel identification, and scenario classification. Due to its meticulously optimized architecture, MTCA is capable of effectively capturing latent associations among different tasks and fully exploiting intrinsic relationships within both the temporal and antenna domains, thereby significantly enhancing overall processing performance.

\subsection{GRU Encoder}
The encoder within our framework is implemented using a GRU. The sequence of complex CSI $\{ H_t \} $ with dimensions $ N_a \times N_t \times N_s $ is initially transformed into a real-valued 2D tensor of size $( N_a \times N_t ) \times ( 2 \times N_s ) $. Here, $ N_a $ represents the number of antennas, $ N_t $ represents the number of time steps, $ N_s $ indicates the number of subcarriers, and 2 corresponds to the real and imaginary parts. Subsequently, the encoder generates the feature $ F \in \mathbb{R}^{(N_a \times N_t) \times (2 \times N_s)} $, which jointly captures both the time domain and the antenna domain.

\subsection{Task-Specified Head}
The MTCA is designed to perform channel prediction, antenna-domain channel extrapolation, channel identification, and scenario classification simultaneously.

\subsubsection{Channel Prediction and Extrapolation}
The CSI prediction problem can be formulated as a sequence-to-sequence prediction task, where the goal is to predict unknown CSI based on known CSI. Specifically, given a sequence of known downlink CSI spanning $ P $ time steps (or antennas), denoted by $ \mathbf{H}_{\text{known}} = \{H_{t-P+1},\ldots,H_{t}\} $, the objective is to predict an unknown downlink CSI sequence spanning $ L $ future time steps (or antennas), represented by $ \mathbf{H}_{\text{unknown}} = \{H_{t+1},\ldots,H_{t+L}\} $. For channel prediction, $ P $ and $ L $ represent 90 and 10, respectively. In the antenna-domain, the corresponding values are 4 and 2. Notably, when channel prediction and antenna-domain extrapolations are performed simultaneously, $ P $ and $ L $ are set to 360 and 20.

To this end, we construct an mapping function $ f_{\Omega} $, which takes the known downlink CSI sequence as input and outputs the predicted unknown CSI sequence:

\begin{equation}\label{eq:capacity}
\hat{\mathbf{H}}_{\text{unknown}} = f_{\Omega}\left(\text{GRUEncoder}(\mathbf{H}_{\text{known}})\right)
\end{equation}

Here, $ f_{\Omega} $ is the FC-based output head designed to capture the mapping relationship between the known CSI sequence features and the unknown CSI, thereby enabling accurate CSI prediction. The model is trained using the Mean Squared Error (MSE) as the loss function, which minimize the error between the predicted and actual CSI sequences.
\begin{equation}\label{eq:mseloss}
\mathcal{L}_{extra} = \frac{1}{N} \sum_{i=1}^{N}  \left\| \mathbf{H}_{unknown} - \hat{\mathbf{H}}_{unknown} \right\|^2 
\end{equation}

\subsubsection{Channel Identification and Scenario Classification}
The channel identification and scenario classification problem can be formulated as a typical classification task, where the goal is to classify the input CSI into categories such as LOS and NLOS, or into more specific scenarios like UMi, UMa, and RMa. This involves binary classification for channel identification and ternary classification for scenario classification.

To achieve this, we construct a mapping function $ f_{\Theta} $, which takes the CSI sequence from the downlink as input and outputs the predicted category $ \hat{y_c} $:

\begin{equation}\label{eq:capacity}
    \hat{y_c} = f_{\Theta}\left(\text{GRUEncoder}(\mathbf{H}_{\text{known}})[:,0]\right),
\end{equation}

Here, $ f_{\Theta} $ is the FC-based output designed to capture the mapping relationship between one of the input CSI sequence features and the corresponding category, thereby enabling accurate identification. The model is trained using the Cross-Entropy Loss as the loss function.
\begin{equation}\label{eq:cross_entropy}
    \mathcal{L}_{nlos(sce)} = -\sum_{c=1}^{C} y_c \log(\hat{y}_c),
\end{equation}
where $ C $ is the number of classes, $ y_c $ is the indicator variable that equals one if class $ c $ is the true class for the sample and zero otherwise, and $ \hat{y}_c $ is the predicted probability that the sample belongs to class $ c $.


\begin{table*}[bp]
\centering
\caption{Simulation Parameter Settings}
\resizebox{0.9\textwidth}{!}{
\begin{tabular}{ 
    >{\centering\arraybackslash}m{7cm} 
    >{\centering\arraybackslash}m{2cm} 
    >{\centering\arraybackslash}m{2cm} 
    >{\centering\arraybackslash}m{2cm} 
}
\hline
\textbf{Parameter}                             & \multicolumn{3}{c}{\textbf{Value}}             \\ \hline
Scenario                                       & UMi            & UMa           & RMa           \\ \cline{2-4} 
Propagation Conditions                         & \multicolumn{3}{c}{LOS, NLOS}                  \\
Center Frequency (GHz)                         & 2              & 2             & 0.7           \\
Bandwidth (MHz)                                & \multicolumn{3}{c}{10}                         \\
Number of Subcarriers                          & \multicolumn{3}{c}{100}                        \\
Initial Number of Clusters                     & \multicolumn{3}{c}{10}                         \\
Number of Sub-paths per Cluster                & \multicolumn{3}{c}{15}                         \\
Transmitter Array Type                         & \multicolumn{3}{c}{UPA} \\
Number of Transmitter Antenna Elements per Row & \multicolumn{3}{c}{2}                          \\
Number of Transmitter Antenna Elements per Column & \multicolumn{3}{c}{4}                                      \\
Spacing between Transmitter Antenna Elements      & \multicolumn{3}{c}{0.5 $\lambda$}             \\
Transmitter Polarization Mode                     & \multicolumn{3}{c}{Omnidirectional $\pm45^\circ$ Dual Polarization} \\
Transmitter Rotation Angle (X-axis)            & \multicolumn{3}{c}{0°}                         \\
Transmitter Rotation Angle (Y-axis)            & \multicolumn{3}{c}{0°}                         \\
Transmitter Rotation Angle (Z-axis)            & \multicolumn{3}{c}{0°}                         \\
Receiver Antenna Polarization                  & \multicolumn{3}{c}{Ideal Point Source}         \\
Transmitter Initial Position (x, y, z) (m, m, m)  & (0, 0, 10)         & (0, 0, 25)        & (0, 0, 35)        \\
Receiver Initial Position (x, y, z) (m, m, m)     & \multicolumn{3}{c}{20, 10, 50}                             \\
Receiver Trajectory Type                       & \multicolumn{3}{c}{Circular}          \\
Receiver Velocity (m/s)                        & \multicolumn{3}{c}{45}                         \\
Movement Duration (s)                          & \multicolumn{3}{c}{20}                         \\
Temporal Sampling Frequency (KHz)              & \multicolumn{3}{c}{1}                          \\ \hline
\end{tabular}
}

\label{tab: parm_setting}

\end{table*}


\subsubsection{Total Loss Function}
The final overall loss $ \mathcal{L} $ is the weighted sum of individual loss functions, as shown in equation~(\ref{eq:loss}):
\begin{equation}
\mathcal{L}_{total} = \sum_{t \in T} \alpha_t L_t,
\label{eq:loss}
\end{equation}
where $ T = \{extra, nlos, sce\} $ represents tasks and $ \alpha_t $ is the weight of task $ t $.The weight for $\mathcal{L}_{extra}$ is set to 9, while the weights for the other components are set to 1.

The higher weight assigned to $\mathcal{L}_{extra}$ reflects its critical role in predicting unknown CSI, which directly impacts the system's ability to adapt and perform in dynamic environments. This weighting strategy ensures that the model prioritizes accurate extrapolation during training, thereby enhancing its predictive capabilities. On the other hand, channel identification and scenario classification tasks, represented by $\mathcal{L}_{nlos(sce)}$, are assigned lower weights due to their relatively lower complexity and more straightforward optimization landscape. This approach helps balance the contributions of different tasks to the overall loss, ensuring that the model achieves a good trade-off between performance on all tasks.

\addtolength{\topmargin}{0.05in}
\section{Simulation results}
In this section, we initially delineate the parameters and configurations of the simulation environment. Subsequently, utilizing the generated data to train the models and evaluate the performance of the proposed MTCA.

\subsection{Dataset}

We used the Southeast University-Purple Mountain Laboratories-6G Pervasive Channel Simulator (SEU-PML-6GPCS) \cite{wang2022pervasive} to simulate the UAV ground-to-air communication channel dataset, based on 3GPP TR36.777\cite{36.777}. We simulated the UAV channel under multiple combinations of scenarios and propagation conditions and obtained the CSI. The parameter settings are shown in Table~\ref{tab: parm_setting}, and the other channel model parameters were set according to the initial recommended values of the simulator. The transmitter is a fixed base station, and the receiver is a moving UAV that travels along a circular trajectory at a speed of 45 m/s.

The configuration of the transmitter base station $ T_x $ includes a UPA with $ N_a = 8 $ antenna elements, arranged with 2 elements per row and 4 elements per column. The spacing between the antenna elements is set to half the wavelength, $ 0.5 \lambda $. The antenna elements are omnidirectional with $ \pm 45^\circ $ dual polarization.

The total number of time steps in the temporal domain is determined by multiplying the movement duration by the temporal sampling frequency, resulting in 20,000 time steps. Sampling is conducted at regular intervals with a stride of 2. This continuous temporal observation facilitates the capture of instantaneous variations and dynamic processes in the channel.

\begin{figure}[tbp]
\centering\includegraphics[width=0.5\textwidth]{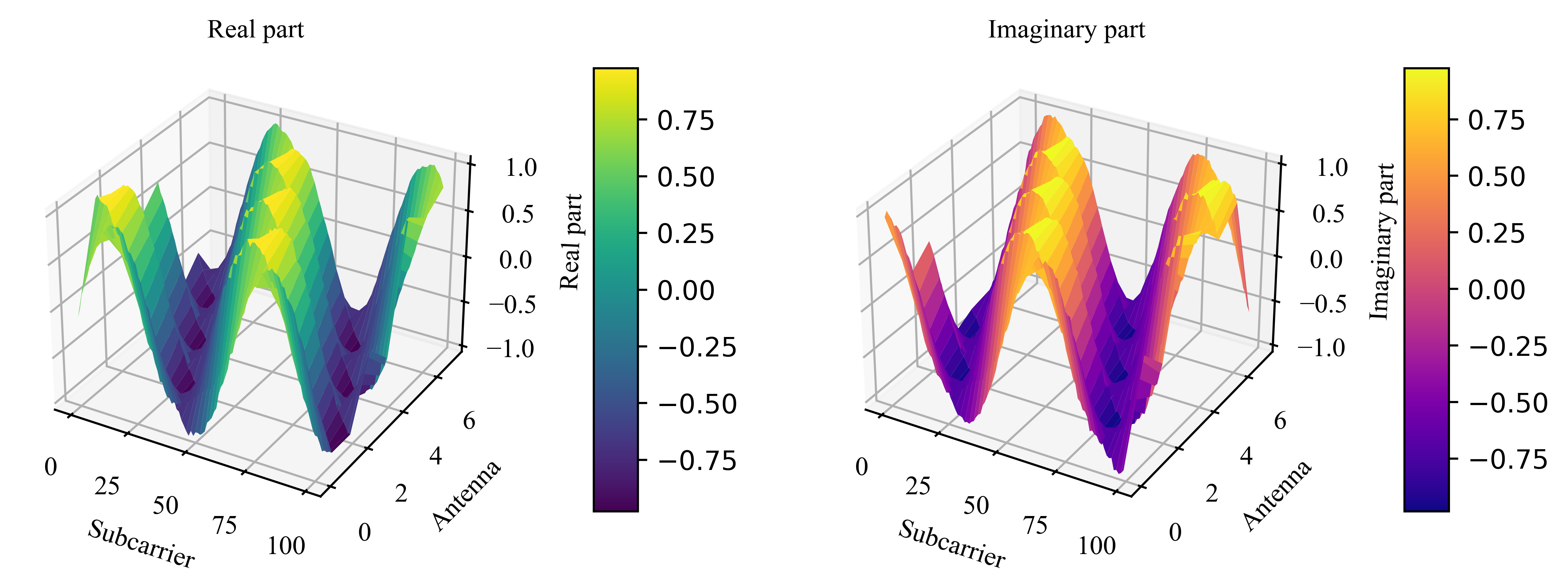}
\caption{The real and imaginary parts of antenna-frequency domain CSI in UMi NLOS scenarios.}
\label{fig:sample}
\end{figure}

For each combination of propagation scenarios, the simulated data were preprocessed and subsequently divided randomly into training and test sets at an 8:2 ratio. As a result, the final dataset comprises 23,765 training samples and 5,941 test samples. An illustrative sample of the real and imaginary parts of antenna-frequency domain CSI under UMi and NLOS conditions is provided in Figure~\ref{fig:sample}.

\begin{table*}[bp]
\centering
\caption{Performance comparison of models under different task combinations.}
\resizebox{0.8\textwidth}{!}{
\begin{tabular}{ccccc}
\toprule
\begin{tabular}[c]{@{}c@{}}Channel Prediction\\ (MSE)\end{tabular} &
  \begin{tabular}[c]{@{}c@{}}Antenna-domain Extrapolation\\ (MSE)\end{tabular} &
  \begin{tabular}[c]{@{}c@{}}Channel Identification\\ (ACC)\end{tabular} &
  \begin{tabular}[c]{@{}c@{}}Scenario Classification\\ (ACC)\end{tabular} &
  \begin{tabular}[c]{@{}c@{}}\#params\\ (M)\end{tabular} \\ \midrule
$3.99\times 10^{-3}$          & /                & /              & /              & 8.6593 \\
$2.37\times 10^{-3}$          & /                & 92.88\%        & /              & 8.6596 \\
$2.56\times 10^{-3}$          & /                & /              & 89.35\%        & 8.6598 \\
$2.06\times 10^{-3}$          & /                & 93.34\%        & 90.88\%        & 8.6602 \\
/                & 0.129            & /              & /              & 8.6584 \\
/                & 0.096            & 91.79\%        & /              & 8.6588 \\
/                & 0.100            & /              & 88.15\%        & 8.6590 \\
/                & 0.092            & 91.27\%        & 88.38\%        & 8.6594 \\
$1.54\times 10^{-3}$          & $1.65\times 10^{-3}$          & /              & /              & 8.6653 \\
$\mathbf{1.43} \times 10^{-3}$ & $\mathbf{1.54} \times 10^{-3}$ & $\mathbf{100\%}$ & $\mathbf{100\%}$ & 8.6662 \\ \bottomrule
\end{tabular}
}
\label{tab:performance_comparison}
\end{table*}

\subsection{Implementation details}
The GRU encoder consists of 6 hidden layers, each with a dimension of 512. Model training and testing were conducted on an NVIDIA GeForce RTX 4090 GPU. The batch size was set to 1024, and the model was trained for a total of 300 epochs. The AdamW optimizer was utilized with a learning rate of 0.0012, which was adjusted using a StepLR scheduler that decays the learning rate by 0.5 every 30 epochs.

\subsection{Results analysis}

As shown in Table \ref{tab:performance_comparison}, a comprehensive comparison of model performance across various task combinations is presented. It is evident that single-task models, whether focusing solely on channel prediction or antenna-domain extrapolation, demonstrate the least satisfactory performance. When channel identification or scenario classification tasks are incorporated, there is a notable reduction in prediction errors. In particular, when channel prediction and antenna-domain extrapolation are performed concurrently, the prediction error shows a remarkable improvement, with almost a 98\% gain in performance for antenna-domain extrapolation. This indicates that MTL can effectively leverage the correlations among different tasks to better extract input features.

Furthermore, when all four tasks - channel prediction, antenna-domain extrapolation, channel identification, and scenario classification - are executed simultaneously, the performance of all tasks is significantly enhanced. Specifically, the accuracy for channel identification and scenario classification reaches 100\%, which further illustrates that the MTCA can better utilize the correlations between the temporal and antenna domains, leading to substantial improvements in wireless communication. It is worth emphasizing that these improvements are achieved with almost no increase in the number of parameters, demonstrating that the proposed MTL architecture not only enhances model performance but also maintains high computational efficiency.

\subsection{Comparison with previous results}

\begin{table}[hbp]
\centering
\caption{Comparative experiments with previous works in single task}

\resizebox{\columnwidth}{!}{
\begin{tabular}{@{}ccc@{}}
\toprule
                      & Channel Prediction & Antenna-domain Extrapolation \\ \midrule
LSTM\cite{Jiang2}                  & $3.35\times10^{-3}$           & 0.1007                       \\
GRU \cite{stenhammar2024comparison}            & $3.99\times10^{-3}$ & 0.1290                       \\
Transformer\cite{Transformer}           & $3.52\times10^{-3}$           & 0.1075                       \\
Transformer-RPE\cite{seq2seq}       & $2.82\times10^{-3}$           & 0.0042                       \\
Seq2Seq-attn-R\cite{seq2seq}        & $1.79\times10^{-3}$           & 0.0033                       \\ \midrule
\textbf{MTCA (Ours)} & $\mathbf{1.43} \times  10^{-3} $ & $\mathbf{0.0015}$              \\ \bottomrule
\end{tabular}
}
\label{tab:constrast}
\end{table}

Table~\ref{tab:constrast} presents the results of the comparative experiments conducted with previous single-task approaches for channel prediction and antenna-domain extrapolation on our proposed dataset. A variety of representative methods were selected as baselines for this comparison, including Long Short-Term Memory (LSTM)\cite{Jiang2}, Transformer\cite{Transformer}, Transformer-RPE\cite{seq2seq}, and Seq2Seq-attn-R\cite{seq2seq}.

LSTM\cite{Jiang2} and GRU\cite{stenhammar2024comparison}, as traditional recurrent neural networks, have been widely used in sequence prediction tasks. LSTM has three gates, which gives it strong memory capabilities. GRU has a simpler architecture with only two gates, making it more efficient in terms of training speed and computational resources. It performs well in capturing short-term dependencies. However, in the context of channel prediction and antenna-domain extrapolation, both LSTM and GRU show certain limitations and relatively inferior performance compared to some more advanced models.

The Transformer\cite{Transformer} has been applied to channel-related predictions nowadays. It can capture long-range dependencies effectively. However, it performs worse in our experiments. Maybe the permutation-invariant in self-attention mechanism and the postional encoding designed for natural language disrupt the sequence information strongly related to position, especially for channel prediction and antenna-domin extrapolation tasks, which inherently contain sequence information from the CSI input \cite{zeng2023transformers}.

Transformer-RPE and Seq2Seq-attn-R\cite{seq2seq} are improved versions based on the original Transformer model. Transformer-RPE introduces the Reverse Positional Encoding (RPE) technique to enhance the robustness of the Transformer. For Seq2Seq-attn-R, the encoder outputs are reversed before applying the attention mechanism, ensuring that the model can effectively handle sequences of varying lengths. From the results, these improvements have led to a significant boost in performance, especially in antenna-domain extrapolation.

However, our proposed MTCA outperforms all these baseline methods in both channel prediction and antenna-domain extrapolation tasks. Specifically, in channel prediction, MTCA achieves a result of $1.43 \times 10^{-3}$ , which is much lower than the other methods. And in antenna-domain extrapolation, its performance is also the best with a value of 0.0015, with an improvement of 54.5\%. This clearly demonstrates that MTCA has a stronger capability to learn joint features between the temporal and antenna domains. By effectively capturing the correlations and patterns in these two domains, it provides more accurate predictions and extrapolations, thus offering substantial support and advantages for downstream tasks. The superior performance of MTCA further highlights its potential and effectiveness in dealing with complex channel-related problems compared to the previous single-task approaches.

\section{Conclusions}
The multi-task learning architecture proposed in this paper, MTCA (Multi-Task Channel Analysis), provides an efficient and comprehensive solution for handling multiple interrelated tasks in wireless communication systems. By sharing feature representations across the temporal and antenna domains, MTCA significantly improves the performance of individual tasks in complex environments. Experimental results indicate that, compared to traditional single-task processing methods, MTCA exhibits stronger learning capabilities and higher adaptability when dealing with diverse CSI.


\bibliographystyle{IEEEtran}
\balance
\bibliography{bibfile}

\end{document}